\documentclass[procedia]{easychair}

\usepackage{doc}
\usepackage{makeidx}
\usepackage{epstopdf}
\def\procediaConference{99th Conference on Topics of
  Superb Significance (COOL 2014)}

\title{ High-temperature expansion for frustrated magnets:
Application to the $J_1$-$J_2$ model on the BCC lattice}

\titlerunning{Thermodynamics of frustrated magnets}

\author{
    Johannes Richter\inst{1}
\and
    Patrick M\"uller\inst{1}
\and
    Andre Lohmann\inst{1}
\and
    Heinz-J\"urgen Schmidt\inst{2}
}

\institute{
Institute for Theoretical Physics, University Magdeburg, 39016
Magdeburg, Germany
  \email{Johannes.Richter@physik.uni-magdeburg.de}
\and
University Osnabr\"uck, Department of Physics, Germany\\
 }

\authorrunning{Richter, M\"uller, Lohmann and Schmidt}

\begin{document}

\maketitle


\begin{abstract}
We present the high-temperature expansion up to 11th order for the
specific heat $C$ and the
uniform susceptibility $\chi_0$ and  up to 9th order for the structure
factor $S_{\bf Q}$ of the frustrated spin-half $J_1$-$J_2$ Heisenberg model on
the BCC lattice. We consider ferromagnetic as well as antiferromagnetic
nearest-neighbor  exchange $J_1$
and frustrating antiferromagnetic  next-nearest-neighbor exchange $J_2$.
We discuss the influence of frustration on the temperature dependence of
these quantities. Furthermore, we use the HTE series  to determine the critical temperature
$T_c$ as a function of the frustration parameter $J_2$.

\end{abstract}


%
%

\section{Introduction}
\label{sect:introduction}

Magnetic systems with strong frustration are currently in the focus of
active theoretical and experimental research \cite{book_Int_to_FrM,our_review}. $J_1$-$J_2$  
Heisenberg models, i.e., models with competing
nearest-neighbor (NN) exchange coupling $J_1$ and next nearest-neighbor (NNN)
exchange coupling $J_2$ can serve as canonical systems to study the interplay of
quantum effects, thermal fluctuations and frustration. On bipartite lattices, such as the square or the BCC
lattices,  the strength of frustration $J_2/J_1$ can be continuously tuned
(where  the limits $J_2=0$ and $J_2 \to \infty$ represent frustration-free
systems), thus allowing to study frustration driven effects in some detail.
The corresponding $J_1$-$J_2$
Heisenberg Hamiltonian is given by
\begin{equation}\label{hamiltonian}
  H=J_1\sum_{\langle i,j\rangle} {\bf S}_i \cdot {\bf S}_j+J_2\sum_{[i,j]} {\bf
 S}_i\cdot {\bf S}_j ,
\end{equation}
where $({\bf S}_i)^2=s(s+1)$,  and $\langle i,j\rangle$ denotes NN
and $[i,j]$ denotes NNN bonds. For antiferromagnetic (AFM) NNN bonds,
$J_2>0$, the spin system is frustrated irrespective of the sign of
$J_1$. The frustration present in the model makes the theoretical treatment
of the model challenging.
There are several investigations on the ground state properties of
two-dimensional (2D) \cite{Sir:2006,Schm:2006,darradi08,ED40,fprg,balents2012,verstrate2013,becca2013,gong2014,eggert2014,shannon2006,richter2010,momoi2011,cabra} and 
three-dimensional (3D)
\cite{bcc_schmidt2002,bcc_oitmaa2004,bcc_majumdar2009,bcc_momoi2013,bcc_RGM2015} 
$J_1$-$J_2$ model.
From these studies it became evident that the 2D  and 3D  models behave
differently.
From numerous studies of the 2D model with AFM $J_1$ it is now clear
that there is an intermediate non-magnetic quantum phase separating
the two semiclassical magnetically ordered ground-state phases, see e.g.
Refs.~\cite{darradi08,ED40,fprg,balents2012,verstrate2013,becca2013,gong2014,eggert2014}. However for
ferromagnetic (FM)
$J_1$ there is a controversial discussion on the existence of such
a non-classical intermediate phase
\cite{shannon2006,richter2010,momoi2011,cabra}.
On the other hand, for the corresponding  3D  BCC model, there are strong
arguments that there is a direct first-order transition at $J_2 = J_2^c$ between the two
magnetically ordered phases present for small (i.e.,  $J_2 < J_2^c$) and  large              
 (i.e.,  $J_2 > J_2^c$) values of $J_2$
 \cite{bcc_schmidt2002,bcc_oitmaa2004,bcc_majumdar2009,bcc_RGM2015}. 
Much less studied are the finite-temperature properties of these models.
In case of strong frustration the quantum Monte Carlo approach is not
applicable.\cite{qmc_sign1}
Therefore, reliable theoretical data for strongly frustrated
quantum spin systems are notoriously rare.
Bearing in mind the very  
active experimental research in the field of frustrated quantum magnetism
\cite{book_Int_to_FrM} theoretical methods to calculate thermodynamic
properties of frustrated magnets are highly desirable.

\begin{figure}[ht!]
	\begin{centering}
	\includegraphics[width=\textwidth,scale=0.95]{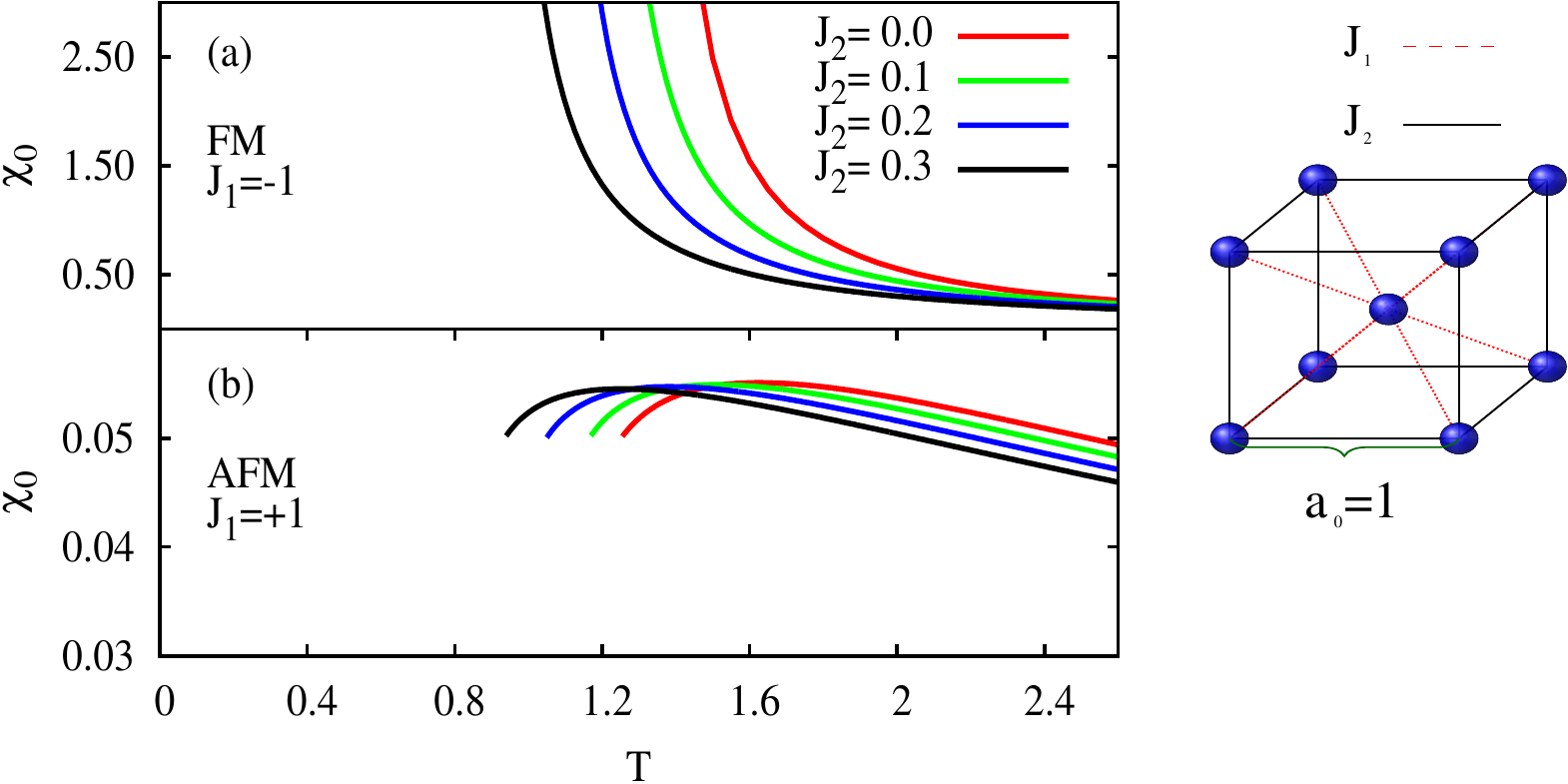}
	\caption{Left: Uniform susceptibility $\chi_0$ as a function of the temperature
	$T$ in the FM (a) and the AFM (b) regime for
	$|J_1|=1$ and different values of the frustrating interaction $J_2$.
	Right: Illustration of the $J_1$-$J_2$ model on the BCC
	lattice.}
	\label{fig:chi0}
	\end{centering}
\end{figure}

\begin{figure}[ht!]
	\begin{centering}
	\includegraphics[scale=0.95]{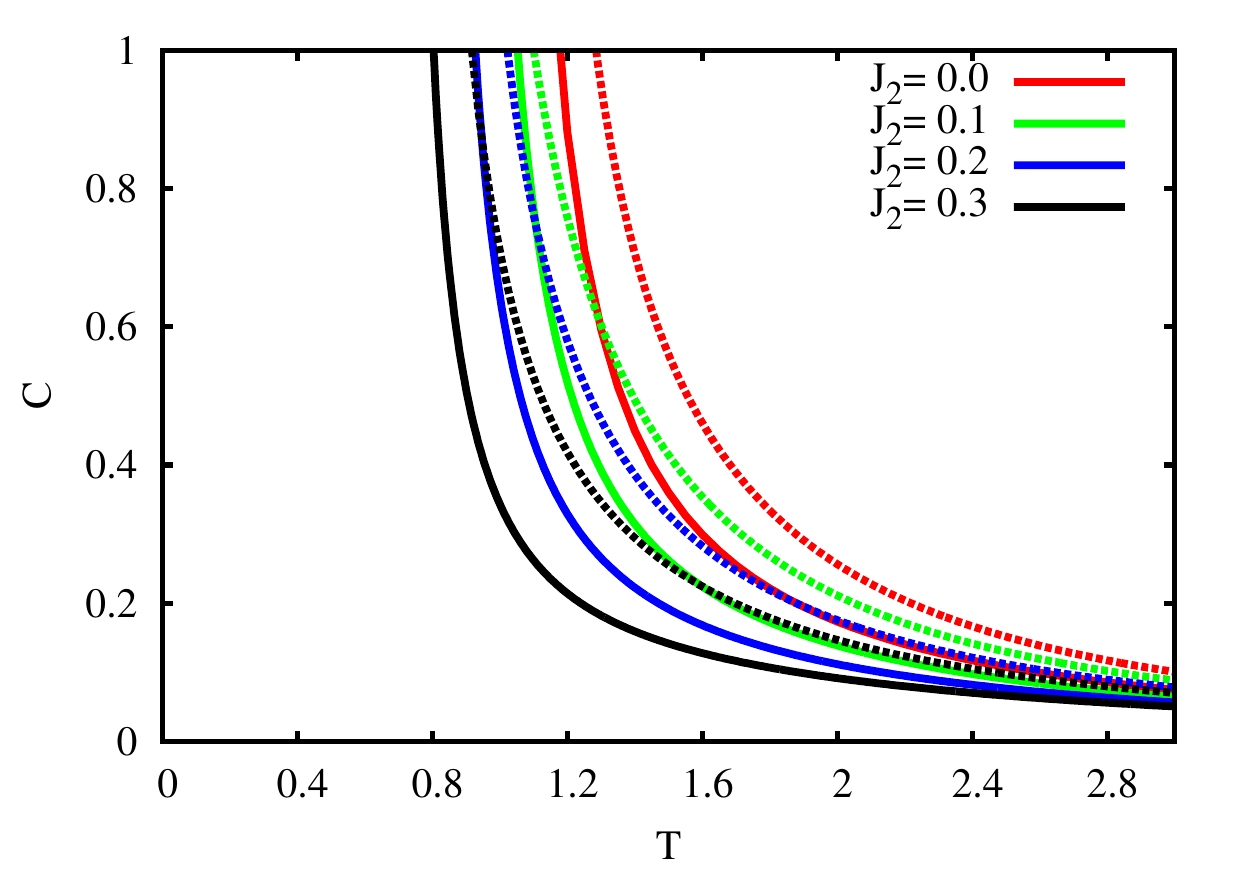}
	\caption{Specific heat $C$ as a function of temperature $T$ in the FM
	($J_1=-1$, solid lines) and the AFM ($J_1=+1$, dashed lines) 
       regime for different values of $J_2$.}
	\label{fig:cV}
	\end{centering}
\end{figure}

A universally applicable method to determine thermodynamic
quantities for magnetic systems (including frustrated ones) is the high-temperature
expansion (HTE).  
In 1950ies and 1960ies this method was 
 developed  and then widely
applied to various Heisenberg magnets, see
Ref.~\cite{domb_green} and references therein.
 An enormous progress could be
achieved in the application of the HTE over the last 20 years by using computer
algebraic tools, see e.g.
\cite{elstner1994,HTE8O_Lohm_11,singh2012,bernu2013,HTE_ten_2014,bernu2015}.
Very recently the present authors have
published a HTE algorithm (encoded as a C++ program available at URL {\tt    
http://www.uni-magdeburg.de/jschulen/HTE/}) to calculate the HTE series of the uniform susceptibility and the
specific heat for
general spin-$s$ Heisenberg models up to 10th order. 
In
\cite{HTE_ten_2014}
we have demonstrated that the susceptibility and the specific-heat data obtained
from the 10th order HTE 
 can be used to
discuss thermodynamic
properties of Heisenberg systems down to moderate temperatures of
about  $T/s(s+1) \sim 0.4 \ldots 0.5 J$.
 
In the present paper we apply our HTE algorithm to investigate the thermodynamics of the $J_1$-$J_2$    
Heisenberg model, see Eq.~(\ref{hamiltonian}), on the BCC lattice and we compare the models with AFM NNN
and FM NN $J_1$. A sketch of the model is given in the right
panel of Fig.~\ref{fig:chi0}.
We focus on spin quantum number $s=1/2$. 
In 3D models there is typically a conventional phase
transition from the high-temperature 
paramagnetic phase with magnetic short-range order to the low-temperature phase with
magnetic long-range order.
We focus here on the paramagnetic regime, i.e. $T \ge T_C$, which is
accessible by HTE, and we discuss the
influence of the frustration on the main thermodynamic quantities, such
as 
 the uniform susceptibility $\chi_0(T)$, the specific heat $C(T)$,
the structure factor $S_{\bf Q}(T)$ and (last but not
least) the critical temperature $T_c$, where the phase transition takes place.

\section{Uniform susceptibility  and specific heat}
\label{C_and_chi}
We start with the discussion of the uniform susceptibility
$\chi_0$ and the specific heat $C$.
For that we can use the HTE algorithm presented in
\cite{HTE_ten_2014}. For the specific lattice at hand we have extended
the HTE to order eleven. It is well-known 
that Pad\'e approximants of the HTE series 
extend the
region of validity of the HTE series down to lower temperatures $T$
\cite{domb_green}.
The Pad\'e 
approximants in form of ratios of two
polynomials $[m,n]=P_m(x)/R_n(x)$ of degree $m$ and $n$ provide an
analytic continuation of a function $f(x)$ given by a power series.
Since typically approximants with $m \sim n$ provide best results,
we use here either the $[5,6]$ or the $[6,5]$ approximants, where  
we choose that approximant that does not have unphysical poles for temperatures in the
region of interest.
The Pad\'e approximant $[6,5]$ for the uniform susceptibility
$\chi_0$ is shown in Fig.~\ref{fig:chi0}.
For FM $J_1=-1$, we find the characteristic divergence of
$\chi_0$ at a critical temperature (Curie temperature)  $T_c$. We will use the critical behavior
in Sec.~\ref{sect:Tc} to determine $T_c$ as a function of $J_2$.
From Fig.~\ref{fig:chi0}(a) it is already obvious that the transition is
shifted to lower values of $T$.
For AFM $J_1=+1$ the susceptibility $\chi_0$ exhibits a
maximum. This maximum  is  related to the critical (N\'{e}el) temperature.
However, the HTE is not able to reproduce the expected kinklike shape of the maximum in
$\chi_0$.
Hence the position $T_{\rm max}$ may yield only a crude estimate of the
N\'{e}el temperature, see Sec.~\ref{sect:Tc}. Nevertheless, the effect of the frustrating bond $J_2$ is
clearly seen.
To get a more reliable determination of the N\'{e}el temperature by a HTE series only the divergence
of the staggered susceptibility, or, alternatively, of the structure factor is
appropriate, see Ref.~\cite{bcc_oitmaa2004,OZ04} and also the next section.
    
Next we discuss the specific heat $C$. It is known that $C$ exhibits
a cusplike singularity at
$T_c$, but no divergence, see, e.g.,
\cite{Wosnitza1988,Landau1991,Janke1993}.
The Pad\'e approximant $[5,6]$ for $C$ 
is shown in Fig.~\ref{fig:cV}.
For both regimes, $J_1=-1$ and $J_1=+1$, specific heat behaves quite similar.
The drastic upturn 
corresponds to  the expected cusplike singularity. 
This upturn takes place at higher temperatures for  AFM $J_1$
than
for FM $J_1$, thus yielding a first indication that the N\'{e}el temperature is larger
than the Curie temperature \cite{OZ04,RGM_layered}.

\begin{figure}[ht!]
	\begin{centering}
	\includegraphics[scale=0.95]{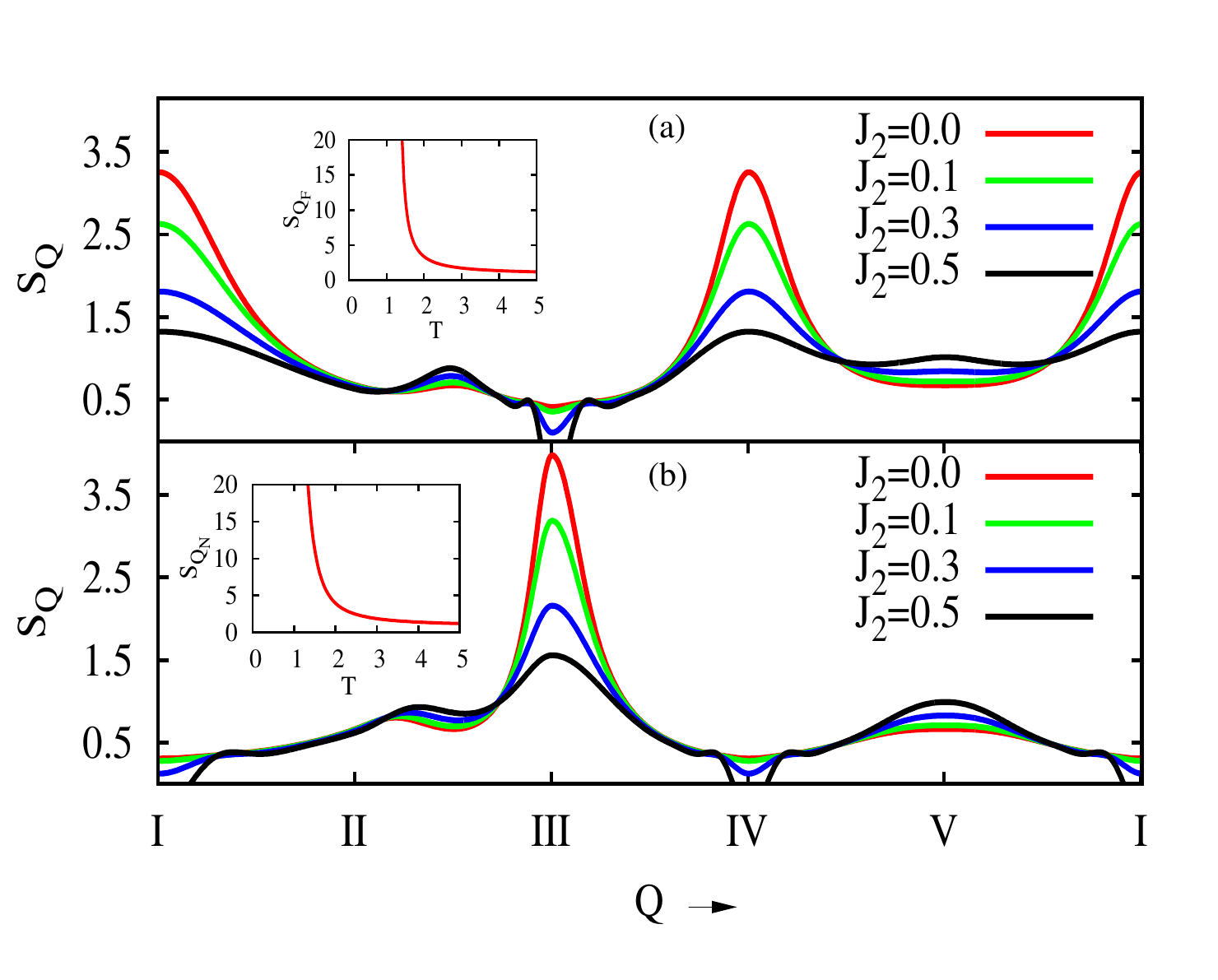}
	\caption{Structure factor $S_{\bf Q}$ as a function of the
	wave vector
	${\bf Q}=(Q_x,Q_y,Q_z)$ for different values of $J_2$
	along several paths between characteristic points (I,II, III, IV,V) in the Brillouin zone (I$=(0,0,0)$; II$=(0,0,\pi)$; III$=(2\pi,2\pi,2\pi)$; IV$=(2\pi,2\pi,0)$; V$=(\pi,\pi,\pi)$). The temperature is set
	to  $T=2$.  Inset: Height of the the maximum in $S_{\bf Q}$ at the
	corresponding magnetic wave vector ${\bf Q}={\bf Q}_M$  in dependence on $T$ for $J_2=0$.
	(a): FM regime ($J_1=-1$), 
	${\bf Q}_M={\bf Q}_F=(0,0,0)$. (b): AFM regime ($J_1=+1$), 
	${\bf Q}_M={\bf Q}_N=(2\pi,2\pi,2\pi)$.
}
	\label{fig:Sq}
	\end{centering}
\end{figure}

\begin{figure}[ht!]
	\begin{centering}
	\includegraphics[scale=0.95]{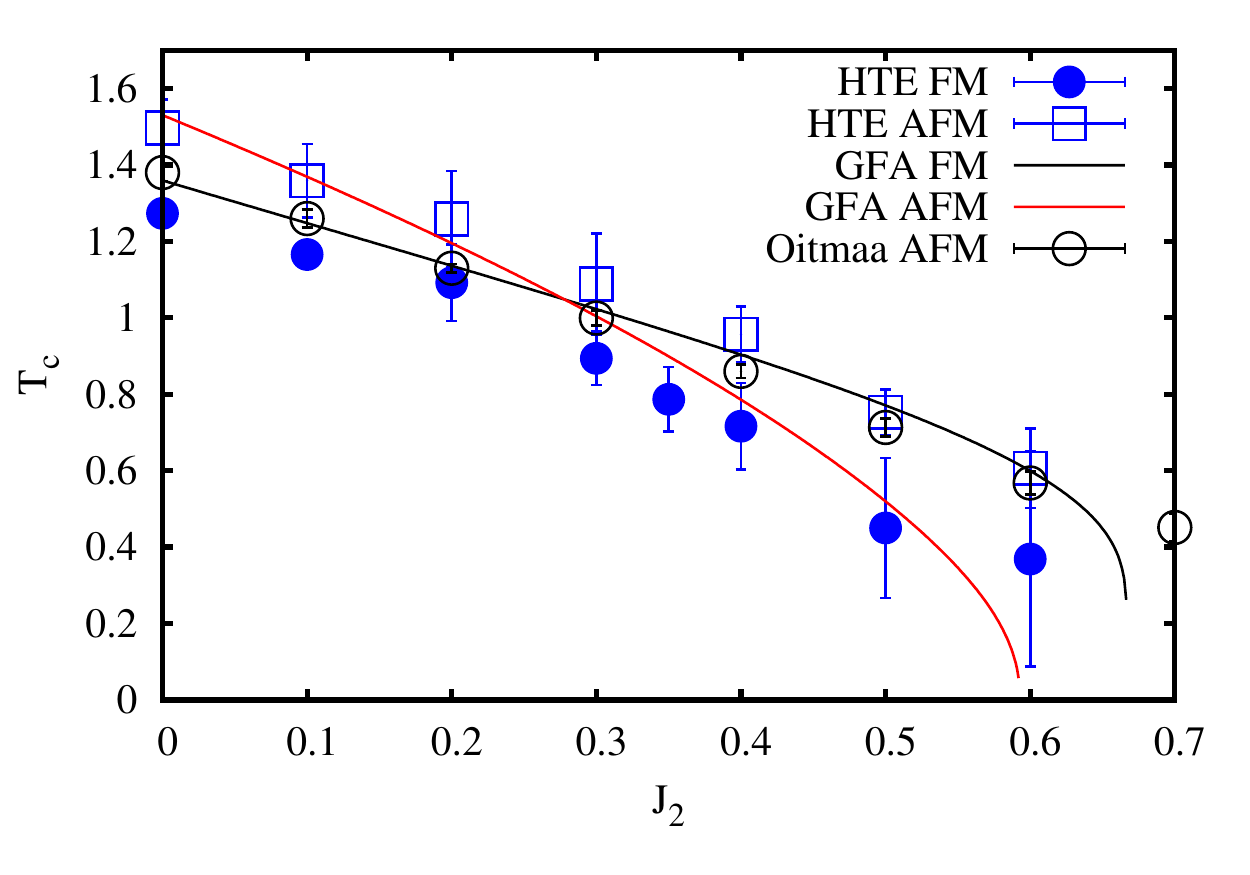}
	\caption{Critical temperatures $T_c$ as a function of the frustrating
	NN coupling $J_2$. Our data are labeled by 'HTE AFM' and 'HTE FM',
	the other labels correspond to data taken from
	\cite{bcc_oitmaa2004} ('Oitmaa AFM'), \cite{bcc_RGM2015}  ('GFA FM') and
	\cite{RGM_inprepar}  ('GFA AFM'). The labels 'FM and 'AFM' correspond
	to $J_1=-1$ and $J_1=+1$, respectively.}
	\label{fig:Tc}
	\end{centering}
\end{figure}




\section{Structure factor}
\label{sect:generalities}
As already mentioned above a reliable determination of the N\'{e}el temperature
requires a HTE series of the staggered susceptibility or the structure
factor at the magnetic wave vector of the N\'{e}el state $\mathbf{Q}_N$.
Both quantities, the staggered susceptibility,  $\chi_{\mathbf{Q}_{N}}$, and the
structure
factor, $S_{\mathbf{Q}_{N}}$,  behave quite similarly.
Approaching a critical point from above, i.e.  
 $T\rightarrow T_{c}^{+}$, one  has $\chi_{\mathbf{Q}_N} \sim \beta S_{\mathbf{Q}_N}$
 \cite{PhysRevB.2.4552}.
Since the  staggered susceptibility is not directly accessible in
experiments we prefer here to use the structure factor  defined by   
\begin{equation}
S_{\mathbf{Q}}=\frac{1}{N}\sum_{n,m}\textrm{cos}(\mathbf{Q}(\mathbf{r}_{n}-\mathbf{r}_{m}))
\langle S_{n}^{z}S_{m}^{z}\rangle
\end{equation}
which can be measured by neutron-scattering experiments.
We use here as basis vectors  of the direct lattice
${\bf a}_{i}=\frac{1}{2}(-1,1,1);
\frac{1}{2}(1,-1,1); 
\frac{1}{2}(1,1,-1)$, cf. the right
panel of Fig.~\ref{fig:chi0}, and  of the reciprocal lattice ${\bf b}_{i}= 2\pi (0,1,1);
2\pi(1,0,1) ; 
2\pi(1,1,0)$. 
Then the
magnetic wave vector ${\bf Q}_M$ of FM state is
${\bf Q}_M=\mathbf{Q}_F=(0,0,0)=(2\pi,2\pi,0)$ and of the AFM N\'{e}el state
is
${\bf Q}_M= \mathbf{Q}_{N}=(2\pi,2\pi,2\pi)$, cf.
Ref.~\cite{bcc_momoi2013}.
 To get $S_{\mathbf{Q}}$  we have extended our HTE code to calculate the spin-spin
correlation functions $\langle S_{n}^{z}S_{m}^{z}\rangle$ entering $S_{\mathbf{Q}}$
up to 9th order (that corresponds to 10th order for the susceptibility).

We show our 9th-order HTE data for $S_{\mathbf{Q}}$ at $T=2$ in Fig.~\ref{fig:Sq}.
The well-pronounced maximum at the corresponding magnetic wave vector ${\bf Q}_M$
is clearly seen. An increase of $J_2$ leads to a suppression of the
maximum, thus indicating the weakening of magnetic short-range order by
frustration. The divergence of $S_{\mathbf{Q}_M}$ as approaching the critical
temperature is shown in the insets of  Figs.~\ref{fig:Sq}(a) and (b).        
Another interesting feature of $S_{\mathbf{Q}}$ is also visible in
Figs.~\ref{fig:Sq}(a) and (b): Although  the frustration leads to a decrease
of  $S_{\mathbf{Q}_M}$, we find an increase of  $S_{\mathbf{Q}}$ at
$\mathbf{Q}=\mathbf{Q}_C=(\pi,\pi,\pi)$.
This increase of $S_{\mathbf{Q}_C}$ with growing $J_2$ is a precursor of the
 so-called collinear AFM phase present for $J_2 > J_2^c$. 
From previous
studies \cite{bcc_schmidt2002,bcc_oitmaa2004,bcc_majumdar2009,bcc_RGM2015}  
it is known that for  $T=0$ the critical $J_2^c$ for  $s=1/2$ is close  its
 classical value $J_2^{c,{\rm clas}} =2/3$, i.e. the maximum
 frustration $J_2=0.5$  used in Fig.~\ref{fig:Sq} is still noticeably below
 $J_2^c$.

\section{The critical temperature $T_c$}
\label{sect:Tc}

We use the HTE series for the structure factor  $S_{\mathbf{Q}}$ 
at $\mathbf{Q}=\mathbf{Q}_M$
to calculate the critical temperature $T_c$.
Assuming critical behavior of the form  $S_{\mathbf{Q}_M} \propto
(T-T_C)^{-\varepsilon}$,
we can use  the well-elaborated technique of the so-called differential
approximants (DA) to extract $T_c$ from the HTE series
of $S_{\mathbf{Q}_M}$, for details see
\cite{bcc_RGM2015,DA3,GUTTMANN2}.
In Fig.~\ref{fig:Tc} we compare our results  with data
obtained by a second-order Green's function
approach (GFA) \cite{bcc_RGM2015,RGM_inprepar} as well as data obtained by
analyzing the HTE series
of the staggered susceptibility \cite{bcc_oitmaa2004}.
We notice that all approaches yield similar results.  There is a significant
reduction of the  critical temperature $T_c$ by frustration. 
Likely, and as
indicated by the GFA data, $T_c(J_2)$ goes to zero as $J_2$ approaches  $J_2^c$.
On the other hand, there is a noticeable difference between the GFA and HTE
data, whereas our HTE data obtained from the structure factor agree
reasonably well with those of \cite{bcc_oitmaa2004} obtained from the staggered susceptibility.  
From a previous comparison between HTE and Monte-Carlo data for unfrustrated
systems \cite{HTE_ten_2014}  we may expect that in the regime of weak
frustration the HTE data for $T_c$ are more accurate than the
GFA data.
Another point concerns the comparison of the Curie temperature $T_C$ with the
N\'{e}el temperature $T_N$. For unfrustrated Heisenberg magnets one finds the
relation $T_N >
T_C$ \cite{OZ04,RGM_layered}.
By contrast to the GFA results, we find that our HTE data obey this relation
in the entire region up to  $J_2 \sim 0.6 $ accessible by the HTE approach.
We mention that the position $T_{\rm max}$ of the 
susceptibility maximum 
for AFM $J_1=+1$, cf. Fig.~\ref{fig:chi0}(a) in Sect.\ref{C_and_chi},
is indeed only about 10\% larger as our $T_N$ values presented in
Fig.~\ref{fig:Tc}.

\section{Summary}

We have applied the high-temperature expansion (HTE) in high orders to investigate 
the thermodynamic quantities 
of frustrated  spin-half $J_1$-$J_2$ Heisenberg
 model on the BCC lattice in the short-range ordered phase at $T \ge T_c$.
We consider FM as well as AFM
NN  exchange $J_1$.
The main focus is on the influence of the frustrating AFM NNN
coupling $J_2$ on the specific heat, the uniform susceptibility, the
structure factor as well as the critical temperature $T_c$.  
The presented  data can be used
to get information on the ratio $J_2/J_1$, e.g., from susceptibility
measurements. Moreover,
our data for the critical temperature as a function of $J_2$  
also provide an information on this ratio.
Interestingly,  the wave-vector dependence of  the  structure factor, although
calculated for $T>T_c$,  
shows some indications of the zero-temperature quantum phase transition at $J_2=J_2^c$, if $J_2$
becomes sufficiently close to $J_2^c$. 
The present investigations are focused on theoretical aspects, however, 
there might be some relevance for FM
compounds \cite{exp2,exp3}.

\label{sect:bib}
\bibliographystyle{unsrt}
\bibliography{richter.bib}

\appendix


\end{document}